\documentclass[twocolumn,english,superscriptaddress,pra]{revtex4-1}

\usepackage{graphicx}% Include figure files
\usepackage{bm}% bold math
\usepackage{amsmath}
\usepackage{amsfonts}
\usepackage{bbm} %bold numbers
\usepackage[FIGTOPCAP]{subfigure}
\usepackage{xr} %allow cross-referencing across files
\usepackage{times}
\usepackage{color}
\usepackage[colorlinks=true,citecolor=blue]{hyperref}

\usepackage{physics}

\newcommand{\dash}{\textemdash}

\newcommand{\vac}{0}
\newcommand{\sect}[1]{\emph{{#1}}.\dash}
\newtheorem{theorem}{Theorem}
\usepackage[capitalise,compress]{cleveref}
\crefname{section}{Sec.}{Secs.}
\crefname{section}{Section}{Sections}
\crefrangelabelformat{equation}{\textup{(#3#1#4)}--\textup{(#5#2#6)}}

\makeatletter
\newcommand*{\addFileDependency}[1]{% argument=file name and extension
  \typeout{(#1)}
  \@addtofilelist{#1}
  \IfFileExists{#1}{}{\typeout{No file #1.}}
}
\makeatother
\newcommand*{\myexternaldocument}[1]{%
    \externaldocument{#1}%
    \addFileDependency{#1.tex}%
    \addFileDependency{#1.aux}%
}
\myexternaldocument{suppmat} %use section labels/equations from suppmat.tex
\newcommand{\J}{\mathcal{J}}

\newcommand{\Lam}{\Lambda}
\newcommand{\al}{\alpha}
\renewcommand{\O}[1]{\mathcal{O}\left(#1\right)}
\newcommand{\W}[1]{\Omega\left(#1\right)}

\makeatletter
\newcommand\footnoteref[1]{\protected@xdef\@thefnmark{\ref{#1}}\@footnotemark}
\makeatother

\begin{document}

\title{Signaling and Scrambling with Strongly Long-Range Interactions}

\author{Andrew~Y.~Guo}
\affiliation{Joint Center for Quantum Information and Computer Science, NIST/University of Maryland, College Park, Maryland 20742, USA}
\affiliation{Joint Quantum Institute, NIST/University of Maryland, College Park, Maryland 20742, USA}
\author{Minh~C.~Tran}
\affiliation{Joint Center for Quantum Information and Computer Science, NIST/University of Maryland, College Park, Maryland 20742, USA}
\affiliation{Joint Quantum Institute, NIST/University of Maryland, College Park, Maryland 20742, USA}
\affiliation{Kavli Institute for Theoretical Physics, University of California, Santa Barbara, CA 93106, USA}
\author{Andrew~M.~Childs}
\affiliation{Joint Center for Quantum Information and Computer Science, NIST/University of Maryland, College Park, Maryland 20742, USA}
\affiliation{Department of Computer Science, University of Maryland, College Park, Maryland 20742, USA}
\affiliation{Institute for Advanced Computer Studies, University of Maryland, College Park, Maryland 20742, USA}
\author{Alexey~V.~Gorshkov}
\affiliation{Joint Center for Quantum Information and Computer Science, NIST/University of Maryland, College Park, Maryland 20742, USA}
\affiliation{Joint Quantum Institute, NIST/University of Maryland, College Park, Maryland 20742, USA}
\author{Zhe-Xuan Gong}
\email{gong@mines.edu}

\affiliation{Department of Physics, Colorado School of Mines, Golden, Colorado 80401, USA}

\date{\today}

\begin{abstract}

Strongly long-range interacting quantum systems\textemdash those with interactions decaying as a power-law $1/r^{\alpha}$ in the distance $r$ on a $D$-dimensional lattice for $\alpha\le D$\textemdash have received significant interest in recent years. They are present in leading experimental platforms for quantum computation and simulation, as well as in theoretical models of quantum information scrambling and fast entanglement creation. Since no notion of locality is expected in such systems, a general understanding of their dynamics is lacking. As a first step towards rectifying this problem, we prove two new Lieb-Robinson-type bounds that constrain the time for signaling and scrambling in strongly long-range interacting systems, for which no tight bounds were previously known. Our first bound applies to systems mappable to free-particle Hamiltonians with long-range hopping, and is saturable for $\alpha\le D/2$. Our second bound pertains to generic long-range interacting spin Hamiltonians, and leads to a tight lower bound for the signaling time to extensive subsets of the system for all $\alpha < D$. This result also lower-bounds the scrambling time, and suggests a path towards achieving a tight scrambling bound that can prove the long-standing fast scrambling conjecture.

\end{abstract}

\maketitle
   
In non-relativistic quantum mechanics, Lieb-Robinson bounds provide a notion of causality \cite{LR}, limiting the speed of information propagation (or signaling) to a finite value in lattice systems with short-range interactions. This bounded signaling speed has strong implications for quantum information and condensed matter physics, leading to entanglement area laws \cite{Hastings07} and the existence of topological order \cite{BravyiHM10}. However, it remains an open question whether the signaling speed must be finite if interactions are long-ranged and decay as an inverse power-law $1/r^{\al}$ in the inter-particle distance $r$. Such power-law interacting systems arise in experimental platforms for quantum computation and quantum simulation, including Rydberg atoms ~\cite{Saffman10}, trapped ions~\cite{Britton12}, polar molecules~\cite{Yan13}, defect centers in solids~\cite{Yao12}, and atoms trapped along photonic crystals~\cite{Douglas15}. The lack of a bounded signaling speed in these systems makes it challenging to understand and predict their dynamics.

For power-law interacting systems with $\al$ greater than the lattice dimension $D$, a finite signaling speed has been shown to persist to some intermediate distance and time \cite{Gong14}. 
At long distances (or times), recent developments show that the signaling speed will diverge at most polynomially in time for $\al >2D$ \cite{Foss-Feig15, Tran18}, ruling out the exponential divergence suggested by earlier results \cite{HK}. 
For $\al\le 2D$, which is the case for most experimental long-range interacting systems, an exponentially growing signaling speed has yet to be ruled out, making the fate of causality far from settled.

In this work, we focus on the regime of \emph{strongly long-range} interacting systems, where interaction energy per site diverges, thus implying $\al\le D$ \cite{Storch15}.
Note that even if one normalizes the interaction strength to make energy \emph{extensive} (i.e. proportional to the number of lattice sites), these systems are still fundamentally different from those with $\al >D$, as energy is in general no longer additive for subsystems \cite{Dauxois}. 
To avoid confusion, we will not perform any normalization of interaction strength throughout this paper, as such normalization can always be performed later by a rescaling of time.

Apart from their existence in experimental platforms \cite{Monroe13, Britton12, Blatt12,Ye13,Lukin17}, strongly long-range interacting systems have received theoretical interest due to their close relation to fast quantum information scramblers \cite{SY93,Bollinger17, Kitaev15, Maldacena16}.
Two fundamental questions about these systems are (1) what is the shortest time $t_{\text{si}}$ needed to send a signal from one site to a site located an extensive distance away, and (2) what is the shortest time $t_{\text{sc}}$ needed to scramble the information stored in the system \footnote{We will define the quantities $t_{\text{si}}$ and $t_{\text{sc}}$ rigorously later in this Letter}?

There have been a number of attempts to answer the above two questions, with limited success. 
For the first question, Refs.~\cite{Eisert13,Hauke13,Eldredge17} show that in certain strongly long-range interacting systems with $\alpha\le D$, information and correlations can spread across the entire system in a finite time, meaning that $t_\text{si}$ has no dependence on the number of sites $N$. 
The Lieb-Robinson-type bound derived in Ref.~\cite{Storch15} does not rule out the possibility of $t_\text{si}$ scaling as $\log(N)\,N^{2\alpha/D}/N^2$ for $\alpha < D$, which vanishes in the $N\rightarrow\infty$ limit \footnote{While the original bound in Ref.~\cite{Storch15} holds for $n$-body interactions, here we cite the bound on the interaction time as it applies to the specific case of two-body interactions.}. However, no protocol that we know of comes close to achieving such fast signaling. 
For the second question, Ref.~\cite{Lashkari13} shows that the scrambling time can be lower-bounded by  $t_\text{sc} \gtrsim 1/N$ for $\alpha=0$, but the fastest known scramblers use a time $t_\text{sc} \propto \log(N)/N$ to scramble the system \cite{SS08}. 

While the ultimate answers to these two questions remain to be found, we present several advances in this Letter. 
First, we prove a new bound for systems that can be mapped to free bosons or fermions with $1/r^{\alpha}$ hopping strength, which leads to $t_\text{si} \gtrsim N^{\alpha/D}/\sqrt{N}$. 
Notably, this bound is tight for $\alpha\le D/2$, as it can be saturated by a new quantum state transfer protocol. 
Second, we prove that $t_{\text{si}}\gtrsim \log(N)N^{\alpha/D}/N$ for general interacting systems, which improves significantly over the previous bound~\cite{Storch15}. 
Building upon this second result, we also prove a tight bound for many-site signaling (from one site to an extensive part of the system). 
This many-site signaling bound leads to a scrambling time bound of $t_\text{sc}\gtrsim N^{\alpha/D}/N$, which generalizes the result in Ref.~\cite{Lashkari13} to all $\alpha<D$. It also reveals a potential way to obtain a tight scrambling bound that can prove the well-known fast scrambling conjecture \cite{SS08}, which remains open for general lattice Hamiltonians, despite recent progress \cite{Lucas18, Lucas19}.

\sect{Tight bound for free particles}We first prove a Lieb-Robinson-type bound for non-interacting bosons/fermions on a lattice. Consider the following free-particle Hamiltonian $H(t)$ defined on a $D$-dimensional lattice $\Lambda$ with $N$ sites:
 \begin{equation}
    \label{eq:freeHam}
 	H(t) = \sum_{\substack{i,j\in\Lambda \\ i< j}} (J_{ij}(t) c_i^\dagger c_j + \text{h.c.}) + \sum_{i\in\Lam}B_{i}(t)c_i^\dagger c_i,
    \end{equation} 
where $c_i^\dagger$ ($c_i$) represents the creation (annihilation) operator. The hopping strength $J_{ij}(t)$ and chemical potential $B_i(t)$ can depend on time and we do not impose any constraint on them for now. We denote an operator $A$ at time $t$ in the Heisenberg picture as $A(t)=U^\dagger(t)A U(t)$, where $U(t) \equiv \mathcal{T} e^{-\frac i \hbar \int_0^t H(t')\,dt'}$ is the time evolution operator ($\hbar = 1$). The operator norm of $A$ will be denoted by $\|A\|$.
\begin{theorem}
    \label{eq:tightbound}
For the Hamiltonian defined in \cref{eq:freeHam} and any pair of distinct sites $X, Y \in \Lam$,
\begin{equation}
    \label{eq:tightboundeq}
    \left\Vert \left[c_{X}(t),c_{Y}^{\dagger}\right]\right\Vert \le  \int_{0}^{t}d\tau\sqrt{\sum_{i\in\Lambda}\left|J_{iX}(\tau)\right|^{2}}.
\end{equation}
\end{theorem}
We use $[\cdot,\cdot]$ to denote the commutator for bosons and anti-commutator for fermions. 

Roughly speaking, the quantity $\|[c_{X}(t),c_{Y}^{\dagger}]\|$ measures the overlap between the support of the operator $c_{X}(t)$ (which expands from site $X$ due to hopping terms) and the site $Y$.
As a result, it also quantifies the amount of information that can be sent between $X$ and $Y$ in a given time $t$. 
Indeed, we define the signaling time $t_\text{si}$ as the minimal time required to achieve $\|[c_{X}(t),c_{Y}^{\dagger}]\| > \delta$ for some positive constant $\delta$. 
Note that we do not expect the chemical potential strength $B_i(t)$ to show up in the bound, as on-site Hamiltonian terms do not change the support of $c_X(t)$.

If the hopping terms in the Hamiltonian are short-ranged (e.g., nearest-neighbor), one might expect $\|[c_{X}(t),c_{Y}^{\dagger}]\|$ to decay exponentially in the distance $r_{XY}$ between $X$ and $Y$, due to the strong notion of causality that follows from the Lieb-Robinson bound \cite{LR}. 
Additionally, if the hopping strength decays as a power law ($|J_{ij}(t)|\le 1/r^{\alpha}$) with $\alpha>D$, intuition would suggest that $\|[c_{X}(t),c_{Y}^{\dagger}]\|$ decays algebraically in $r_{XY}$ \cite{HK,Gong14}, indicating a weak notion of causality. 
However, the right-hand side of \cref{eq:tightboundeq} has no dependence on $r_{XY}$.
This is because the bound is tailored to strongly long-range hoppings with $\alpha<D$, which makes it loose for shorter-ranged long-range hoppings.

Assuming that $|J_{ij}(t)|\le 1/r^{\alpha}$, we can simplify \cref{eq:tightboundeq} to
\begin{equation}
\left\Vert \left[c_{X}(t),c_{Y}^{\dagger}(0)\right]\right\Vert \le t\times\begin{cases}
\O{1} & \al > D/2,\\
\O{N^{\frac{1}{2}-\al/D}} & 0\le \al\le D/2.
\end{cases}\label{eq:corollary}
\end{equation}
where $N$ is the number of lattice sites and $\mathcal{O}$ is the asymptotic ``big-$\mathcal{O}$'' notation \cite{bigO}. Therefore, for $\al\le D/2$, it takes a time $t_\text{si} = \W{N^{\alpha/D}/\sqrt{N}}$ \cite{bigO} to signal from site $X$ to site $Y$, independent of the distance between $X$ and $Y$. 

In the next section, we show that for $\alpha\le D/2$, the bound in \cref{eq:tightboundeq} can be saturated (up to a factor of 2) by engineered free-particle Hamiltonians. 
This leads to the conclusion that causality can completely vanish\dash in the sense that signals can be sent arbitrarily fast given large enough $N$\dash for a strongly long-range hopping system with $\alpha<D/2$.
It remains an open question whether such a statement can be generalized to systems with $D/2 \le\alpha<D$ for either free or interacting particles.

\sect{Proof of \cref{eq:tightbound}}Let us first go into the interaction picture of $\sum_i B_{i}(t)c_i^\dagger c_i$ to eliminate the on-site terms from the Hamiltonian in \cref{eq:freeHam} \footnote{
This imparts a time-dependent phase $e^{i\phi_{jk}(t)}$ onto the hopping term $J_{jk}(t)$ for some $\phi_{jk}(t)\in[0,2\pi)$ and $j\neq k$.
Since this phase does not change the value of $|J_{jk}(t)|$, it does not affect the bound.}. 
We now have a pure hopping Hamiltonian $ H_I(t) = \sum_{ij} \tilde J_{ij}(t)c_i^\dagger c_j$ with $|\tilde J_{ij}(t)|\equiv|J_{ij}(t)|$. Because $ H_I(t)$ is a quadratic Hamiltonian, $c_X(t)$ is a time-dependent linear combination of annihilation operators on every site, and we can write $ [c_X(t), c^\dagger_Y] \equiv f_{XY}(t) \mathbbm{1}$, where $f_{XY}(t)$ is a number and $\mathbbm{1}$ represents the identity operator. Given that $U_I(t)\ket{\vac}=\ket{\vac}$, where $U_I(t)$ is the time-evolution operator corresponding to $H_I(t)$, and $c_X(t)\ket{\vac}=0$, we have 
\begin{equation}
	\label{eq:deffxy}
	f_{XY}(t) = \bra{\vac}[c_X(t), c^\dagger_Y]\ket{\vac} = \bra{\vac} c_X(0)U_I(t)c^\dagger_Y \ket{\vac}.
\end{equation} 
For convenience, we define the (normalized) states $\ket{\psi_X} \equiv c^\dagger_X(0)\ket{\vac}$ and $\ket{\psi_Y(t)} \equiv U_I(t)c_Y^\dagger\ket{\vac}$. Taking the time derivative of \cref{eq:deffxy} gives
\begin{equation}
	\label{eq:onelineproof}
	\frac{df_{XY}}{dt} = -i\bra{\psi_X} H_I(t)\ket{\psi_Y(t)}. 
\end{equation}
By the Cauchy-Schwarz inequality, 
\begin{align}
\left\Vert  \frac{df_{XY}}{dt}\right\Vert &\le \left\Vert H_I(t)\ket{\psi_X} \right \Vert \: \left\Vert \ket{\psi_Y(t)} \right \Vert\\
&=\left\Vert H_I(t)\ket{\psi_X} \right \Vert= \sqrt{\sum_{i\in\Lambda}\left| \tilde J_{iX}(t)\right|^{2}}.
\end{align}
The last equality follows from $\ket{\psi_X}$ being a single excitation localized on site $X$ and $ H_I(t)$ consisting only of hopping terms $\tilde J_{ij}(t)c_i^\dagger c_j$. Applying the fundamental theorem of calculus yields the bound on  $f_{XY}(t)$ and hence \cref{eq:tightbound}.

\begin{figure}
\includegraphics[width=0.9\linewidth]{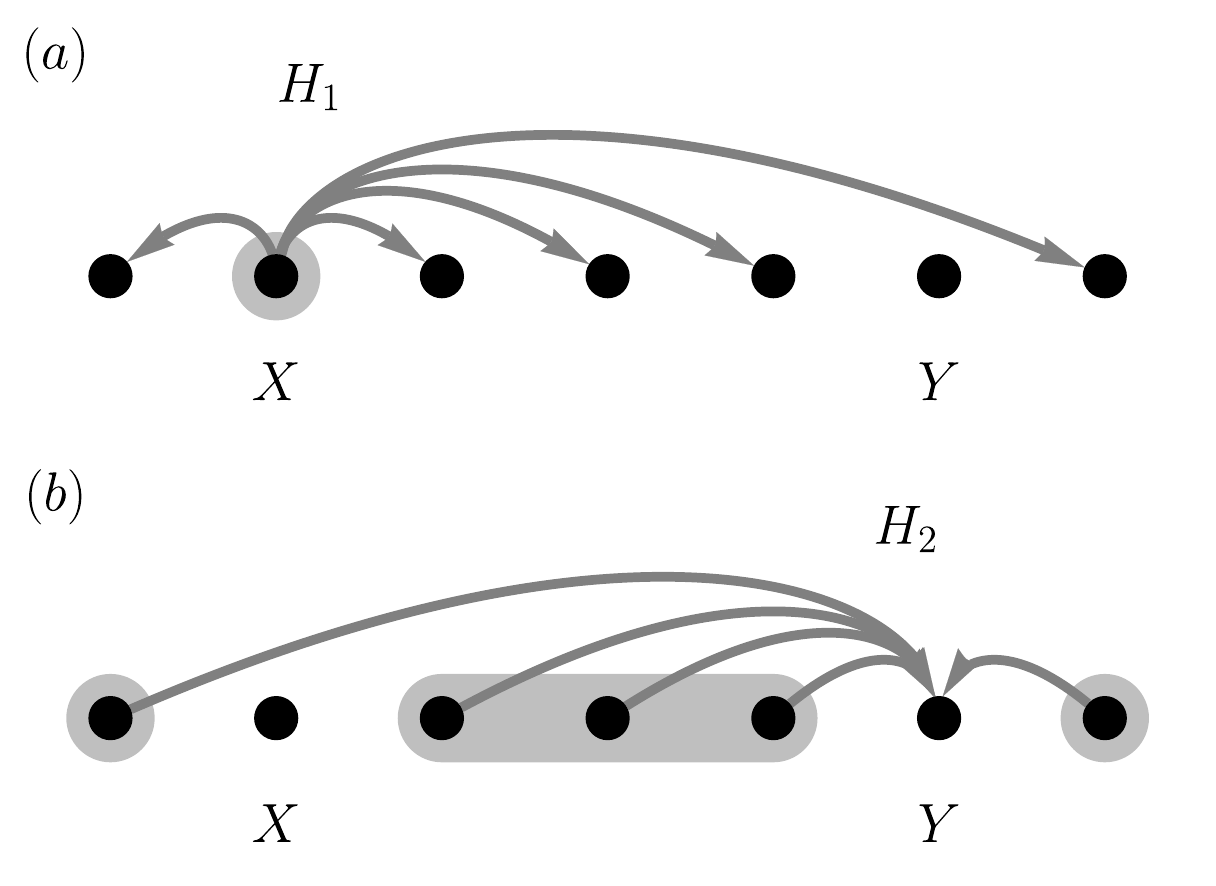}
\caption{A fast quantum state transfer protocol for a long-range Hamiltonian acting on a lattice of dimension $D=1$ with $N=7$ sites. The strengths of the hopping terms are bounded by a power-law $1/r^{\al}$ in the distance $r$. The active interactions in each time-step are depicted as directed edges with uniform weights. (a) The site $X$ is initially in the state $\ket{\psi}$ (gray circle), with the other (unoccupied) sites in state $\ket{0}$. Time-evolving by the Hamiltonian $H_1$ for time $\O{N^{\al/D-1/2}}$  (indicated by gray arrows) yields a superposition of the $\ket{0}^{\otimes N}$ state and a symmetric $\ket{W}$ state over the remaining $N-2$ sites. (b) Applying the Hamiltonian $H_2$ for the same duration of time completes the state transfer of $\ket{\psi}$ to the target site $Y$.}
\label{Fig_Gong}
\end{figure}

\sect{Saturating the free-particle bound}We now show that the bound in \cref{eq:tightbound} can be saturated by engineered Hamiltonians that can also be used to perform fast quantum state transfer. In particular, the protocol presented here has a state transfer time of $T = \O{N^{\alpha/D}/\sqrt{N}}$, which\dash for $\alpha\le D/2$\dash improves over the fastest-known state transfer protocol using long-range interactions \cite{Eldredge17}.

Our setup for the state transfer task is depicted in \cref{Fig_Gong}. We initialize a lattice with $N$ sites in a tensor product of unoccupied states $\ket{0}$ and some unknown normalized bosonic/fermionic state $\ket{\psi} = a\ket{0}+b\ket{1}$ on a single site $X$. 
The goal of state transfer is to move $\ket{\psi}$ to the target site $Y$ after the system time-evolves by a $\ket{\psi}$-independent (but possibly time-dependent) Hamiltonian $H(t)$ \cite{NJ2014,Epstein17}.

The unitary time-evolution operator $U(T)$ can be said to implement state transfer in time $T$ if it satisfies the following condition:
\begin{align}
	\label{eq:fidelity}
	\left|\bra{0}_{X}\bra{0}^{\otimes N-2}\bra{\psi}_{Y} U(T)\ket{\psi}_{X}\ket{0}^{\otimes N-2}\ket{0}_{Y}\right| = 1.
 \end{align}
We refer to the left-hand side of \cref{eq:fidelity} as the \emph{fidelity} of the state transfer, which can be bounded directly by a Lieb-Robinson-type bound on $H(t)$ such as \cref{eq:tightboundeq} \cite{Epstein17}. 

We label the sites that are not $X$ or $Y$ by $1,\dots,N-2$ and denote the furthest distance between any pair of sites by $L=\O{N^{1/D}}$. Our state transfer protocol is given by the following piece-wise time-independent Hamiltonian:
\begin{align}
\label{eq:protocol}
H(t)= \begin{cases}
	H_{1}=\frac{1}{L^{\al}} \sum_{i=1}^{N-2} c_{X}^{\dagger}c_{i}+\text{h.c.} & 0<t<\frac{T}{2},\\
	H_{2}=\frac{1}{L^{\al}} \sum_{i=1}^{N-2} c_{i}^{\dagger}c_{Y}+\text{h.c.} & \frac{T}{2}<t<T,
\end{cases}
\end{align}
where $T=\pi L^{\al}/\sqrt{N-2}$ is the total time for the protocol. Note that while $H(t)$ satisfies the constraint $|J_{ij}(t)|\le 1/r_{ij}^{\alpha}$ assumed in \cref{eq:corollary}, the corresponding $J_{ij}(t)$ terms do not actually vary with the distances between sites.

Evolving the initial state 
$\ket{\Psi} \equiv \ket{\psi}_{X}\ket{0}^{\otimes N-2}\ket{0}_{Y}$  by $H_1$ for time $T/2$ yields the intermediate state 
\begin{equation}
	e^{-i H_1 T/2}\ket{\Psi}= a \ket{0}^{\otimes N} + b\ket{0}_{X}\ket{W}\ket{0}_{Y}.
\end{equation}
Here, $\ket{W}= \frac{1}{\sqrt{N-2}}\sum_{i=1}^{N-2} c_{i}^{\dagger}\ket{0}^{\otimes N-2}$ is the W state over the $N-2$ remaining sites. Further evolving the state by $H_2$ for time $T/2$ yields the final state: 
\begin{equation}
	e^{-i H_2 T/2}e^{-i H_1 T/2}\ket{\Psi}= \ket{0}_{X}\ket{0}^{\otimes N-2}(a \ket{0}_{Y}+ b\ket{1}_{Y}).
\end{equation}
Thus we have achieved perfect quantum state transfer in time $T = \O{N^{\al/D}/\sqrt{N}}$. 
Note that the distance between $X$ and $Y$ on the lattice does not appear in the state transfer time.
    Setting $b=1$ in the above protocol leads to
\begin{equation}
    \expval{[c^\dagger_{X}(T),c_{Y}]}{\Psi} = \frac{1}{2}\int_{0}^{T}d\tau\sqrt{\sum_{i\in\Lambda}\left|J_{iX}(\tau)\right|^{2}}.
\end{equation} 
Thus, the bound in \cref{eq:tightboundeq} is saturated up to a factor of 2.

It should be pointed out that, for $\al > D/2$, the above protocol requires a time that increases with $N$, which is slower than for the previous result in Ref.~\cite{Eldredge17}. 
While that protocol has a state transfer time that is constant for $\alpha\le D$, it uses an engineered Hamiltonian with interactions, and therefore cannot be applied to systems of non-interacting particles.
In general, allowing interactions may increase the rate of information propagation, and proving a Lieb-Robinson-type bound in these situations requires a different approach. 

\sect{Improved bound for general interacting systems}We now derive bounds on the signaling time that extend beyond free-particle Hamiltonians. 
Without loss of generality, we study a generic interacting spin Hamiltonian $ 	H(t) = \sum_{i<j} h_{ij}(t)$ where $\|h_{ij}(t)\|\le 1/r_{ij}^{\alpha}$ and on-site interactions have been eliminated by going into an interaction picture. 
We will bound the quantity $\|[A(t),B]\|$, where $A$ and $B$ are arbitrary operators supported on sets of sites $X$ and $Y$ respectively,  using the following Lieb-Robinson series \cite{HK}:
\begin{align}
	\label{eq:HKseries1}
	\|[A(t),B]\| &\le 2\|A\|\|B\||X||Y|\sum_{k=1}^{\infty}\frac{(2t)^{k}}{k!} \J^k(X,Y), \\
    \label{eq:hoppingterms}
	\J^k(X,Y) &\equiv \sum_{i_{1},\dots,i_{k-1}}J_{Xi_{1}}J_{i_{1}i_{2}}\dots J_{i_{k-1}Y}.
\end{align}
Here, $|X|$ stands for the number of sites $X$ acts on. Each term in \cref{eq:hoppingterms} represents a sequence of $k$ directed hops in the lattice that originates at site $X$ and ends at site $Y$.
For distinct sites $i$ and $j$, $J_{ij}=1/r_{ij}^{\al}$ represents a directed hop from $i$ to $j$. 
We set $J_{ii} = \sum_{j\ne i} J_{ij}$ \footnote{
The strength of the on-site hop $J_{ii}$ is defined this way for technical reasons that we explain in the Supplemental Material.}.

Since $J_{ij}$ decays slowly in $r_{ij}$ for $\alpha<D$, our improved bound on $\|[A(t),B]\|$ requires bounding each term in \cref{eq:hoppingterms} using new techniques beyond those in previous efforts \cite{HK,Storch15} and detailed in the Supplemental Material~\cite{SM}. The result (assuming $\alpha<D$) is:

\begin{equation}
\label{eq:newbound}
	\|[A(t),B]\| \le  2\|A\|\|B\||X||Y|\left(\frac{e^{\Theta(N^{1-\al/D})t}-1}{\Theta(N^{1-\al/D})r_{XY}^\al}\right).
\end{equation}
The factor $\Theta(N^{1-\alpha/D})$ \cite{bigO} comes from the total interaction energy per site given by $J_{ii}$.

We consider now the case of signaling between subsystems $X$ and $Y$ of a system $\Lam$ with $|X|,|Y|=\O{1}$. We formally define $t_\text{si}$\dash the signaling time from $X$ to $Y$\dash as the smallest time $t$ such that there exist unit-norm operators $A$ and $B$ supported on $X$ and $Y$ respectively and a constant $\delta=\Theta(1)$ such that $\|[A(t),B]\| > \delta$ \cite{Lashkari13}. If we further assume that $X$ and $Y$ are separated by an extensive distance $r_{XY}=\W{N^{1/D}}$, then the following lower bound holds for the signaling time: 
\begin{equation}
	\label{eq:tsi}
	t_\text{si}=  \W{\frac{\log(N)}{N^{1-\al/D}}}. 
\end{equation}
While we do not know of any examples that saturate this bound, it is the tightest-known signaling time bound for strongly long-range interacting systems.
Indeed, the bound is close to being saturated in the limit of $\alpha\rightarrow D^-$, as the state transfer protocol in Ref.~\cite{Eldredge17} shows that $t_\text{si}=\O{\log(N)}$ can be achieved at $\alpha=D$. Unfortunately, generalizing our bound in \cref{eq:newbound} to the case of $\alpha=D$ leads to $t_\text{si}=\W{1}$ \cite{SM}, which is not saturated by Ref.~\cite{Eldredge17}.

\sect{Many-site signaling and scrambling bounds}Of recent interest in the fields of theoretical high-energy and condensed matter physics is the phenomenon of quantum information scrambling \cite{Lashkari13,Swingle16,SS08,Hayden07,Xu18}. 
Previous work on scrambling in power-law interacting systems has focused primarily on numerical analysis \cite{Pappalardi18,Zhou18}, whereas general mathematical results are lacking. 
Only in all-to-all interacting systems (which can be treated as the limit $\al=0$) have Lieb-Robinson-type bounds been used to bound the scrambling time \cite{Lashkari13}. 
Using the bound derived in \cref{eq:newbound}, we can prove a scrambling time bound for systems with $0<\alpha<D$, a regime for which no better result is known. 

To derive a bound on the scrambling time, we first derive a bound on the many-site signaling time. We define the many-site signaling time $t_\text{ms}$ to be the smallest $t$ required to signal from $X$ to a $Y$ that has extensive size.
Lieb-Robinson-type bounds such as \cref{eq:newbound} naturally limit the time for many-site signaling. However, a direct application of \cref{eq:newbound} to many-site signaling leads to a loose bound. Instead, a more refined technique that sums over all sites within the subsets $X$ and $Y$ yields a tighter bound \cite{Gong17}:
\begin{equation}
\label{refinedbound}
    \|[A(t),B]\|\le  2\|A\|\|B\|\sum_{i\in X, j\in Y}\frac{e^{\Theta(N^{1-\al/D})}t-1}{\Theta(N^{1-\al/D})r_{ij}^\al}.
\end{equation}
This bound reduces to \cref{eq:newbound} when $|X|,|Y|=1$.

A system on a lattice $\Lam$ is said to scramble in a time $t_\text{sc}$ if any information initially contained in a finite-sized subsystem $S \subset \Lam$ is no longer recoverable from measurements on $S$ alone \cite{tscdefn}.
That information is not lost, however, but can be recovered from the complement $\bar{S} = \Lam \setminus S$ of $S$ \cite{Hayden07,Cotler18}. 
As a result, scrambling implies the ability to signal from $S$ to $\bar{S}$ \cite{Lashkari13}. 
Thus, $t_\text{sc}$ is lower bounded by the time it takes to signal from a subset $S$ with size $\Theta(1)$ to its complement with size $\Theta(N)$, which corresponds to the many-site signaling time.

Using \cref{refinedbound}, we obtain the following scrambling time bound for $0\le\alpha<D$:
\begin{equation}
    \label{eq:manysitebound}
    t_\text{sc}\ge t_\text{ms}= \W{ \frac{1}{N^{1-\al/D}}}.
\end{equation}
Note that this bound differs from \cref{eq:tsi} by a $\log(N)$ factor.
Additionally, although the bound on $t_\text{si}$ in \cref{eq:tsi} may allow further tightening, the bound on $t_\text{ms}$ in \cref{eq:manysitebound} cannot be generically improved for $0\le\alpha<D$. 
To see this, we consider a long-range Ising Hamiltonian $H=\sum_{i\neq j}J_{ij}\sigma_{i}^{z}\sigma_{j}^{z}$, with $J_{ij}=1/r_{ij}^{\al}$. 
For simplicity, we consider the subset $S$ to be a single site indexed by $i$ and construct operators $A=\sigma_{i}^{+}$ and $B=\bigotimes_{j\ne i}\sigma_{j}^{+}$ that are supported on $S$ and $\bar{S}$ respectively. 
We can analytically calculate the expectation value of $[A(t),B]$ in an initial state $\ket{\psi}=\frac{1}{\sqrt{2}}[\ket{0}^{\otimes N}+\ket{1}^{\otimes N}]$ \cite{Gong17}:
\begin{equation}
    \label{eq:manysiteprotocol}
\bra{\psi}[A(t),B]\ket{\psi}=\sin(2t\sum_{j\ne i}J_{ij}).
\end{equation}
Using $J_{ij}=1/r_{ij}^{\al}$, we find that the signaling time of this protocol is $t=\O{N^{\al/D-1}}$ for $0\le\al<D$, which saturates the many-site signaling time bound in \cref{eq:manysitebound} \footnote{A different protocol for many-site signaling was given in \cite{Eisert13}. That result yields a many-site signaling time of $t_\text{ms} =\O{N^{\al/D-1/2}}$, which does not saturate the bound in \cref{eq:manysitebound}.}.
This does not, however, imply that the corresponding scrambling time bound is tight.
In fact, the fastest known scramblers in long-range interacting systems with $\alpha=0$ have a scrambling time $t_\text{sc}=\O{\log(N)/N}$ \cite{Lashkari13}, which is a factor of $\log(N)$ away from saturating the bound. This suggests that future improvements on the scrambling time bound may be possible.

\sect{Conclusions and Outlook}In this Letter, we make several advances in bounding the signaling and scrambling times in strongly long-range interacting systems.
Our results suggest a number of possible future directions. One is to find the optimal signaling time bound for general strongly long-range interacting systems. Previously, this has been an outstanding challenge; we now know of a free-particle bound that is tight for $\al \in [0,D/2]$ and a general bound that is nearly tight as $\alpha\rightarrow D^-$. The search for the optimal bound for $\alpha\in [0,D]$ has thus been narrowed down significantly.

Additionally, our bound for signaling to an extensive numbers of sites hints at a strategy for achieving a better scrambling bound.
In particular, the protocol that saturates our many-site signaling bound relies on an initial entangled state, whereas the definition of scrambling assumes that the system begins in a product state \cite{tscdefn}. 
It may be possible to improve the scrambling time bound by explicitly restricting our attention, when bounding $t_\text{ms}$, to initial product states. Obtaining an extra $\log(N)$ factor in $t_\text{sc}$ by such restriction would prove the main part of the fast-scrambling conjecture in Ref.~\cite{SS08}.

Finally, we expect that the improved Lieb-Robinson-type bounds derived in this work may lead to a better understanding of the spreading of correlations \cite{Luitz19}, the growth of entanglement entropy \cite{Gong17}, and thermalization timescales \cite{Nandkishore15} in strongly long-range interacting systems. 

\sect{Acknowledgments}
We thank B. Swingle, M. Foss-Feig, J. Garrison, Z. Eldredge, and A. Deshpande for helpful discussions. 
MCT, AYG, and AVG acknowledge funding by the DoE ASCR Quantum Testbed Pathfinder program, DoE BES Materials and Chemical Sciences Research for Quantum Information Science program, NSF Ideas Lab on Quantum Computing, ARO MURI, AFOSR, ARL CDQI, and NSF PFC at JQI. 
This work was supported in part by the Army Research Office (MURI award number W911NF-16-1-0349), the Canadian Institute for Advanced Research, the National Science Foundation (under Grant Nos.~CCF-1813814 and PHY-1748958), the Heising-Simons Foundation, and the U.S. Department of Energy, Office of Science, Office of Advanced Scientific Computing Research, Quantum Algorithms Teams and Quantum Testbed Pathfinder (award number DE-SC0019040) programs.
ZXG acknowledges funding from NSF RAISE-TAQS program.
AYG is supported by the NSF Graduate Research Fellowship Program under Grant No. DGE-1322106.
%We also would like to thank @zeldredge for running Overheard on Quant-Ph. 

\bibliographystyle{apsrev4-1}
\bibliography{LRbib}

\begin{thebibliography}{48}
\bibitem [{big()}]{bigO}%
  \BibitemOpen
  \href@noop {} {}\bibinfo {note} {The ``big-$\mathcal{O}$'' notation $f(x) =
  \O{g(x)}$ indicates the existence of constants $c_1$ and $N_1$ such that
  $0\le f(x) \le c_1g(x)$ for all $x\ge N_1$. We also use the ``big-$\Omega$''
  notation $f(x) = \W{g(x)}$ to indicate that $g(x) = \O{f(x)}$. Finally, we
  use the ``big-$\Theta$'' notation $f(x)=\Theta(g(x))$ when both $f(x)=\O
  {g(x)}$ and $f(x) = \W {g(x)}$.}\BibitemShut {Stop}%
\bibitem [{tsc()}]{tscdefn}%
  \BibitemOpen
  \href@noop {} {}\bibinfo {note} {The definition we use for the scrambling
  time $t_\text{sc}$ is based on the one introduced in Ref.~\cite{Lashkari13}.
  Roughly, it corresponds to the minimal time required for a system of $N$
  spins to evolve from a product state to a state that is nearly maximally
  entangled on all subsystems of size $kN$ for some constant
  $0<k<\frac12$.}\BibitemShut {Stop}%
\end{thebibliography}
\end{document}

\title{Supplemental Material for ``Signaling and Scrambling with Strongly Long-Range Interactions''}

\author{Andrew~Y.~Guo}
\affiliation{Joint Center for Quantum Information and Computer Science, NIST/University of Maryland, College Park, Maryland 20742, USA}
\affiliation{Joint Quantum Institute, NIST/University of Maryland, College Park, Maryland 20742, USA}
\author{Minh~C.~Tran}
\affiliation{Joint Center for Quantum Information and Computer Science, NIST/University of Maryland, College Park, Maryland 20742, USA}
\affiliation{Joint Quantum Institute, NIST/University of Maryland, College Park, Maryland 20742, USA}
\affiliation{Kavli Institute for Theoretical Physics, University of California, Santa Barbara, CA 93106, USA}
\author{Andrew~M.~Childs}
\affiliation{Joint Center for Quantum Information and Computer Science, NIST/University of Maryland, College Park, Maryland 20742, USA}
\affiliation{Department of Computer Science, University of Maryland, College Park, Maryland 20742, USA}
\affiliation{Institute for Advanced Computer Studies, University of Maryland, College Park, Maryland 20742, USA}
\author{Alexey~V.~Gorshkov}
\affiliation{Joint Center for Quantum Information and Computer Science, NIST/University of Maryland, College Park, Maryland 20742, USA}
\affiliation{Joint Quantum Institute, NIST/University of Maryland, College Park, Maryland 20742, USA}
\author{Zhe-Xuan Gong}
\email{gong@mines.edu}
\affiliation{Department of Physics, Colorado School of Mines, Golden, Colorado 80401, USA}

\date{\today}
\maketitle

In \cref{sec:generalbound} of this Supplemental Material, we provide a detailed proof of the general Lieb-Robinson-type bound for long-range interactions with $\al\le D$ mentioned in Eq.~(15) of the main text. The bound has a closed-form expression that can be used to lower bound the signaling time [see Eq.~(16)] for $\al \le D$.

In \cref{sec:exactsumbound}, we derive a second general Lieb-Robinson-type bound that is the tightest one can get from the Lieb-Robinson series mentioned in Eq.~(13), although it lacks a closed-form analytic expression.
We show numerically that the signaling time obtained from this bound has the same scaling as a function of system size $N$ as the bound in Eq.~(15) of the main text when $N$ is sufficiently large. 
Therefore, the bound presented in the main text is\dash in a broad sense\dash the best one can obtain without developing new techniques beyond those used in deriving the traditional Lieb-Robinson series.

\section{Proving the general Lieb-Robinson-type bound for $\alpha\le D$}
\label{sec:generalbound}
Before we present the proof of the general Lieb-Robinson-type bound given in Eq.~(15) of the main text, let us summarize in \cref{sec:math} some mathematical preliminaries useful for the proof. 

\subsection{Mathematical preliminaries}
\label{sec:math}
In this section, we elaborate on the scaling of the on-site hop parameter $J_{ii}$ defined after Eq.~(14). 
We define the quantity $\lambda = \max_{i\in\Lambda} J_{ii}$, which for power-law interactions has strength
\begin{equation}
	\label{eq:self-hop}
	\lam = \max_{i\in\Lam} \sum_{j\in\Lam\backslash i} J_{ij} = \max_{i\in\Lam}\sum_{j\in\Lam\backslash i} \frac{1}{r_{ij}^{\al}}.
\end{equation}
If the lattice $\Lam$ is a square lattice with unit spacings, then $\lambda$ scales as
\begin{equation}
        \label{eq:lamdef2}
	\lam = \begin{cases}
        \Th{N^{1-\al/D}} & \text{for }0\le\al<D,\\
        \Th{\log N} & \text{for }\al=D,\\
        \Th 1 & \text{for }\al>D.
\end{cases}
\end{equation}
In general, the scaling of $\lam$ as a function of $N$ in \cref{eq:lamdef2} holds asymptotically for large regular lattices \cite{Storch15}.

For $\al \le D$, we note that $\lam$ diverges in the thermodynamic limit. 
For some applications, it is preferred to apply a normalizing factor of $1/\lam$ (due to Kac \cite{Kac63}) to the Hamiltonian to ensure the system energy is extensive. Since experimental systems (such as those with dipolar interactions in 3D) do not necessarily have extensive energy, we would prefer not to apply the Kac normalization. The light cone contours for Kac-normalized Hamiltonians follow straightforwardly from our results upon rescaling the time by a factor of $\lambda$. 

In the rest of this subsection, we justify the dependence of $\lambda$ on $N$ and $\al$ as shown in \cref{eq:lamdef2}.
We assume that the lattice $\Lam$ is $D$-dimensional with $N = L^D$ sites. 
Without loss of generality, let $i$ be the site located at the origin and define $r_j \equiv r_{ij} \ge 1$. 
For $0\le \al < D$, we bound \cref{eq:self-hop} above by an integral:
\begin{align}
	\sum_{j\in\Lam} \frac{1}{r_{j}^{\al}} &\le \int_{\vec{r}\in \mathbb{R}^D} \frac{d^D\vec{r}}{\|\vec{r}\|^\al} 
	= \frac{2\pi^{\frac D 2}}{\Gamma(\frac D2)} \int_0^L \frac{dr}{r^{\al-D+1}}  = \frac{\omega_D}{D-\al}L^{D-\al},
\end{align}
where $\omega_D\equiv 2\pi^{\frac D 2}\big/\Gamma(\frac{D}{2})$ is the hyper-surface area of a unit $D$-sphere and $\Gamma(\cdot)$ is the Gamma function.  It follows that $\lam = \O{ N^{1-\al/D}}$  for $\al<D$. 
The asymptotic lower bound $\lam = \W{N^{1-\al/D}}$ follows from setting $\|\vec r\| \rightarrow \|\vec r\| +\sqrt D$ and integrating from $r=1$ to $r=\infty$.

For $\al = D$, we perform the same calculation, taking care to avoid integrating over the origin:
\begin{align}
	\sum_{j\in\Lam} \frac{1}{r_{j}^{\al}} &\le  \int_{\|\vec{r}\| \ge 1} \frac{d^D\vec{r}}{\|\vec{r}\|^D} + \sum_{j \in \Lam} \theta \left( 1+\sqrt{D} - r_j\right)
    \\ & \le  \omega_D \int_1^L \frac{dr}{r} + \frac{\omega_D\left(1+\sqrt D+\frac12\right)^D}{\omega_D\left(\frac12\right)^D}
    \\ &= \frac {\omega_D}D \log(N) + \left(2\sqrt D+3\right)^D,
\end{align}
where $\theta(\cdot)$ denotes the Heaviside step function. So, at the critical point $\al=D$, we have $\lam = \O{ \log(N)}$.
The lower bound $\lam = \W{\log(N)}$ holds in a similar fashion.

For $\al > D$, the sum in \cref{eq:self-hop} converges, so $\lam$ can be bounded by a constant independent of $N$. 
Thus, we have verified the asymptotic scaling of the on-site parameter $\lam$ for the three cases listed in  \cref{eq:lamdef2}.

\subsection{Proof of the bound in Eq.~(15)}
\label{sec:newbound}

In this section, we provide a simple proof of Eq.~(15) for long-range interactions. 
First, let us recall the Lieb-Robinson series from Eq.~(13) of the main text.
\begin{align}
	\label{eq:HKseries2}
		\|[A(t),B]\| &\le 2\|A\|\|B\||X||Y|\sum_{k=1}^{\infty}\frac{(2t)^{k}}{k!} \J^k(X,Y), \\
	\label{eq:hopping2}
	\J^k(X,Y) &\equiv \sum_{i_{1},\dots,i_{k-1}}J_{Xi_{1}}J_{i_{1}i_{2}}\dots J_{i_{k-1}Y}.
\end{align}
We use the so-called \emph{reproducibility condition} for finite systems with power-law decaying interactions \cite{HK, Gong14}. Specifically, for $\al > 0$ and distinct $i,j\in \Lam$, the second-order hopping term $k=2$ in \cref{eq:hopping2} can be bounded by
\begin{equation}
	\label{eq:repcon}
	\J^2(i,j) = \sum_k J_{ik}J_{kj} \le p\lam J_{ij},
\end{equation}
where $p=2^{\al+1}$ and $\lam$ is the on-site hop parameter defined in \cref{eq:self-hop}. 
This inequality allows the power-law decay of $J_{ij}$ to be reproduced across multiple hopping terms. 

We reorder the summations on the right-hand side of \cref{eq:hopping2} by introducing a new index $n$ to count the number of self-hops in a particular sequence of hopping sites $\{i_1,\dots,i_{k-1}\}$.
Specifically, $n$ represents the number of indices $j\in\{0,\dots,k-1\}$ such that $i_{j}=i_{j+1}$ (with $i_{0}=X$ and $i_{k}=Y$). 
Now let us first assume that the $n$ self-hops occur in the first $n$ terms ($J_{Xi_1}$ to $J_{i_{n-1}i_{n}}$). Then we may rewrite the right-hand side of \cref{eq:hopping2} as
\begin{equation}
\label{eq:selfhopresum}
\sum_{i_{1},\dots,i_{k-1}}J_{Xi_{1}}\dots J_{i_{k-1}Y} 
= \lambda^n \sum_{i_{n+1},\dots,i_{k-1}}J_{i_{n}i_{n+1}}\dots J_{i_{k-1}Y},
\end{equation}
using the fact that each self-hop term $J_{ii}$ is equal to $\lam$. 
If, on the other hand, the $n$ self-hops appear in arbitrary positions in the sequence of hops, then accounting for these cases multiplies \cref{eq:selfhopresum} by the combinatorial factor of $\binom k n$.
Inserting into \cref{eq:hopping2} gives
\begin{align}
	\label{eq:choose}
	\J^k(X,Y) = \sum_{n=0}^{k}{\binom k n}\lambda^{n} \left[\sum_{i_{n+1},\dots,i_{k-1}}J_{i_Xi_{n+1}}\dots J_{i_{k-1}Y} \right],
\end{align}
where we relabeled $i_n$ as $X$. 
Now, using the fact that $i_{j}\neq i_{j+1}$ for $j=n,\dots,k-1$ (where $i_k=Y$), we can apply the reproducibility condition in \cref{eq:repcon} a total of $k-n$ times along with the normalization condition $J_{ij} = 1/r_{ij}^{\al}$ for $i\neq j$ to get
\begin{align}
	\sum_{i_{n+1},\dots,i_{k-1}}J_{Xi_{n+1}}\dots J_{i_{k-1}Y} \le (\lam p)^{k-n-1} J_{XY} = \frac{(\lam p)^{k-n-1}}{r_{XY}^{\al}}.
\end{align}
Finally, inserting this inequality into \cref{eq:choose} and applying the binomial theorem gives
\begin{equation}
\mathcal{J}^{k}(r_{XY}) \le \sum_{n=0}^{k}{\binom k n}\lambda^{n}\left[\frac{(\lambda p)^{k-n-1}}{r_{XY}^{\al}}\right]\ = \frac{(\lambda+\lambda p)^{k}}{\lambda pr_{XY}^{\al}}.
\end{equation}
Inserting this bound for $\mathcal{J}^{k}(r_{XY})$ into \cref{eq:HKseries2} gives an exponential series bound for the commutator
\begin{align}
	\|[A(t),B]\| &\le 2\|A\|\|B\||X||Y|\sum_{k=1}^{\infty}\frac{(2t)^{k}}{k!} \J^k(X,Y) \\
    &\le 2\|A\|\|B\||X||Y|\sum_{k=1}^{\infty}\frac{(2t)^{k}}{k!} \frac{(\lambda(1+p))^{k}}{\lambda pr_{XY}^{\al}}\\
    &= 2\|A\|\|B\||X||Y| \left(\frac{e^{2\lam t(1+p)}-1}{\lambda pr_{XY}^{\al}}\right) \\
    &= 2\|A\|\|B\||X||Y|\left(\frac{e^{\Theta(N^{1-\al/D})t}-1}{\Theta(N^{1-\al/D})r_{XY}^\al}\right), \label{eq:newbound2}
\end{align}
which reproduces Eq.~(15) in the main text.

\section{Second general Lieb-Robinson-type bound for $\al \le D$}\label{sec:exactsumbound}
The above derivation of the bound in \cref{eq:newbound2} requires the use of the reproducibility condition~[\cref{eq:repcon}], which can make the bound loose compared to the Lieb-Robinson series in \cref{eq:HKseries2}. In this section, we will exactly sum the series in \cref{eq:HKseries2} and compare the resulting bound to that of \cref{eq:newbound2}. We will show by numerical analysis that\dash although the bound obtained directly from the Lieb-Robinson series is tighter than the one \cref{eq:newbound2}\dash it largely shares the same scaling behavior when the number of sites $N$ is large.

\subsection{Summing the Lieb-Robinson series exactly}
\label{sec:exactsumboundproof}
We now exactly calculate the sum in \cref{eq:HKseries2} without using the reproducibility condition. Since \cref{eq:HKseries2} is an infinite series, one cannot perform the sum directly. But using a discrete Fourier transform, the series can be summed numerically in a highly efficient manner.

To use the discrete Fourier transform, we assume that $J_{ij}=1/r_{ij}^\al$ is translationally invariant. Note that the physical interaction strength between lattice sites does not need to be translationally invariant, as it just needs to be bounded by $J_{ij}$. his additional assumption does not greatly affect the generality of the following results.

For simplicity, we consider a 1D lattice $\Lam$ that consists of $N$ spins on a ring; the following results generalize straightforwardly to lattices in arbitrary dimensions. Let $r_{ij} = \min\{|i-j|,|N-i+j|\}$ be the (periodic) distance metric, which coincides with the graph distance $d(i,j)$ on $\Lam$. Due to translational invariance, we denote $J_{ij}$ by $J(r_{ij})$ which satisfies $J(r+N) = J(r)$.

We now perform a discrete Fourier transform from the position space parameterized by $r=0,1,2,\cdots,N-1$ to a momentum space parameterized by $p=0,1,2,\cdots,N-1$, denoted by $\mathcal{F}_p[f(r)]=\sum_{r=0}^{N-1}e^{-2\pi ipr/N}f(r)$. 
We observe that the sum in the definition of $\J^k(X,Y)$ in \cref{eq:hopping2} can be rewritten as the $k$-fold convolution of $J(r)$ with itself. 
Thus, letting $r\equiv r_{XY}$ and $\J^k(r) \equiv J^k(X,Y)$, the discrete Fourier transform of \cref{eq:hopping2} is given by
\begin{align}
	\mathcal{F}_p[\J^{k}(r)]  = \omega(p)^k,
\end{align}
where $\omega(p)=\mathcal{F}_p[J(r)]$.
We now take the discrete Fourier transform of the entire series in \cref{eq:HKseries2}:
\begin{align}
\mathcal{F}_p\left[	\|[A(t),B]\|\right] & \le\sum_{k=1}^{\infty}\frac{(2t)^{k}}{k!}\omega(p)^k = e^{2\omega(p)t}-1. \label{eq:fcomm}
\end{align}
The series in \cref{eq:HKseries2} can thus be evaluated exactly by taking the inverse Fourier transform of \cref{eq:fcomm}:
\begin{equation}
	\label{eq:exactbound}
		\|[A(t),B]\| \le \mathcal{F}_r^{-1}\left[e^{2\omega(p) t}-1\right].
\end{equation}
where $\mathcal{F}_r^{-1}[g(p)]\equiv\frac{1}{N}\sum_{p=0}^{N-1}e^{2\pi ipr/N}g(p)$ defines the inverse discrete Fourier transform. 
For $\al = 0$, the inverse Fourier transform can be evaluated to yield the analytical expression \begin{equation}
    \|[A(t),B]\| \leq 2\|A\|\|B\||X||Y|\left(\frac{e^{4Nt}-1}{N}\right),
\end{equation}
which matches the bound in \cref{eq:newbound2} up to constant factors. 
For $\al > 0$, it is difficult to obtain a simple analytical expression for the bound in \cref{eq:exactbound}. We will thus evaluate this bound numerically, as detailed in the next section.

\subsection{Numerical comparison of the two bounds}
\label{exactboundnumerics}

In this section, we study the asymptotic scaling of the exact summation bound in \cref{eq:exactbound}.
Because the discrete Fourier transform can be performed rather efficiently using the Fast Fourier Transform algorithm \cite{FFT65} numerically, we can evaluate the right-hand side of \cref{eq:exactbound} for system sizes up to the order of $10^6$. 
This allows us to compare the bound in \cref{eq:exactbound} (referred to below as the "exact summation bound") with the bound in \cref{eq:newbound2} (referred to below as the "analytical bound") for large $N$.

\begin{figure}[h]
	\includegraphics[width=.5\linewidth]{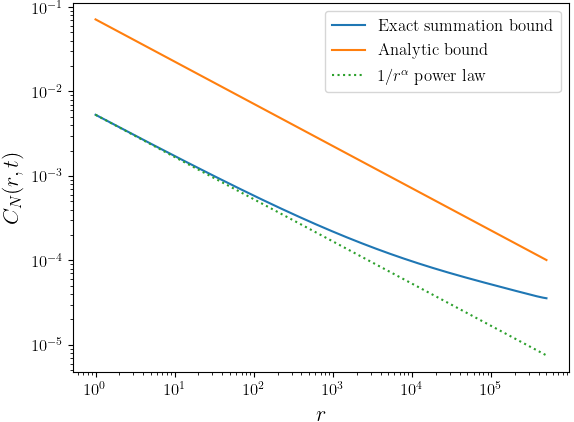}
    \caption{A comparison between the exact summation bound [\cref{eq:exactbound}] and the analytical bound [\cref{eq:newbound2}] sites as a function of the distance $r$ between operators $A$ and $B$. The specific plot assumes a 1D periodic lattice with $N = 10^6$, $\al = 0.5$, $t = 1/\lam$, and $r=1,2,\cdots,N/2$.}
    \label{Fig_Rdep}
\end{figure}

Let us first focus on the large-$r$ asymptotic behavior of the two bounds. 
As a typical example, we set $N = 10^6$ and $t = 1/\lam$ and plot the right-hand side of the two bounds (with $\|A\|=\|B\|=|X|=|Y|=1$) as a function of $r$ for $\al = 0.5$ in \cref{Fig_Rdep}. 
Unsurprisingly, the right-hand side of the analytic bound in \cref{eq:newbound2} decays as $1/r^{\al}$ for the entire range of $r$. 
The exact summation bound leads to the same scaling for small $r$, but not for large $r$. 
While this comparison leaves room for potential tightening of the bound in \cref{eq:newbound2} for large $r$, generic improvement of the bound for all $r$ seems unlikely. 

Next, we compare the $N$-dependence of the two bounds. To get rid of the $r$-dependence, we will compare the signaling times between two sites with either $r=1$ (the smallest possible separation on a 1D ring) or $r=N/2$ (the largest possible separation). 
If the two bounds agree with each other at both $r=1$ and $r=N/2$ in the large $N$ limit, it is reasonable to believe that they will agree with each other at all values of $r$.

For $\al < 1$, the analytical bound gives the following signaling time (upon setting $\|[A(t),B]\|= 1$) as function of $r$ and $N$:  
\begin{equation}
	\label{eq:contour_al<D}
	t_\text{si}(r,N) =\W{\frac{\log(N^{1-\al}r^{\al})}{N^{1-\al}}}.
\end{equation} 
Choosing either $r=1$ or $r=N$ leads to $t_\text{si}=\Omega(N^{\al-1}\log(N))$, consistent with Eq.~(16) of the main text. For the exact summation bound, we numerically compute $t_\text{si}$ by finding the value of $t$ that makes $\|[A(t),B]\| = 1$ over a range of $N$ from $10^4$ to $10^6$ for both $r = 1$ and $r = N/2$. We then fit $t$ as a function of $N$ to the function $a N^\gamma \log(N)$.
\begin{figure*}[h]
	\subfigure[~$r = 1$]{
	\includegraphics[width=.45\textwidth]{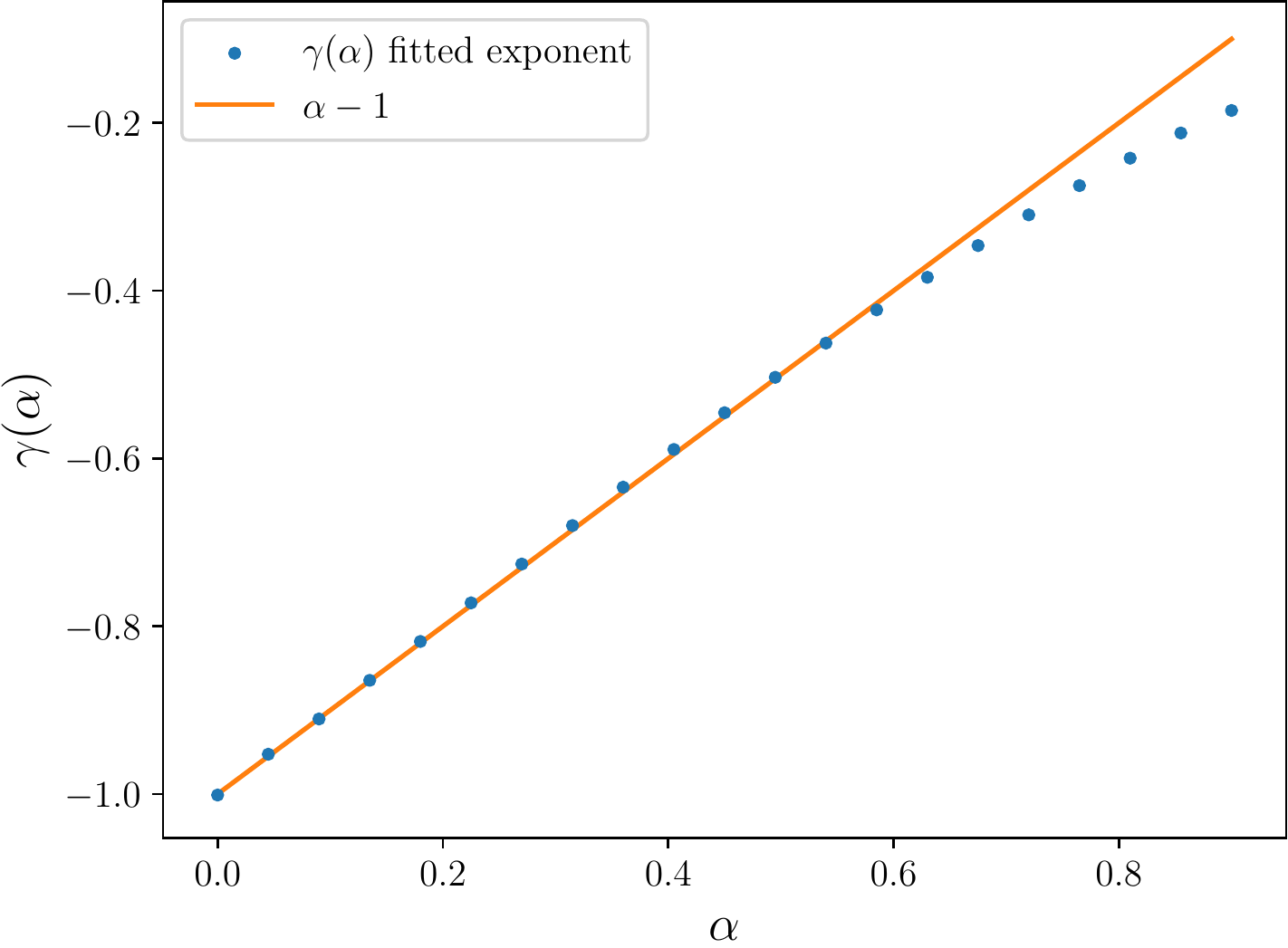}
    } \qquad
    \subfigure[~$r = N/2$]{
    \includegraphics[width=.455\textwidth]{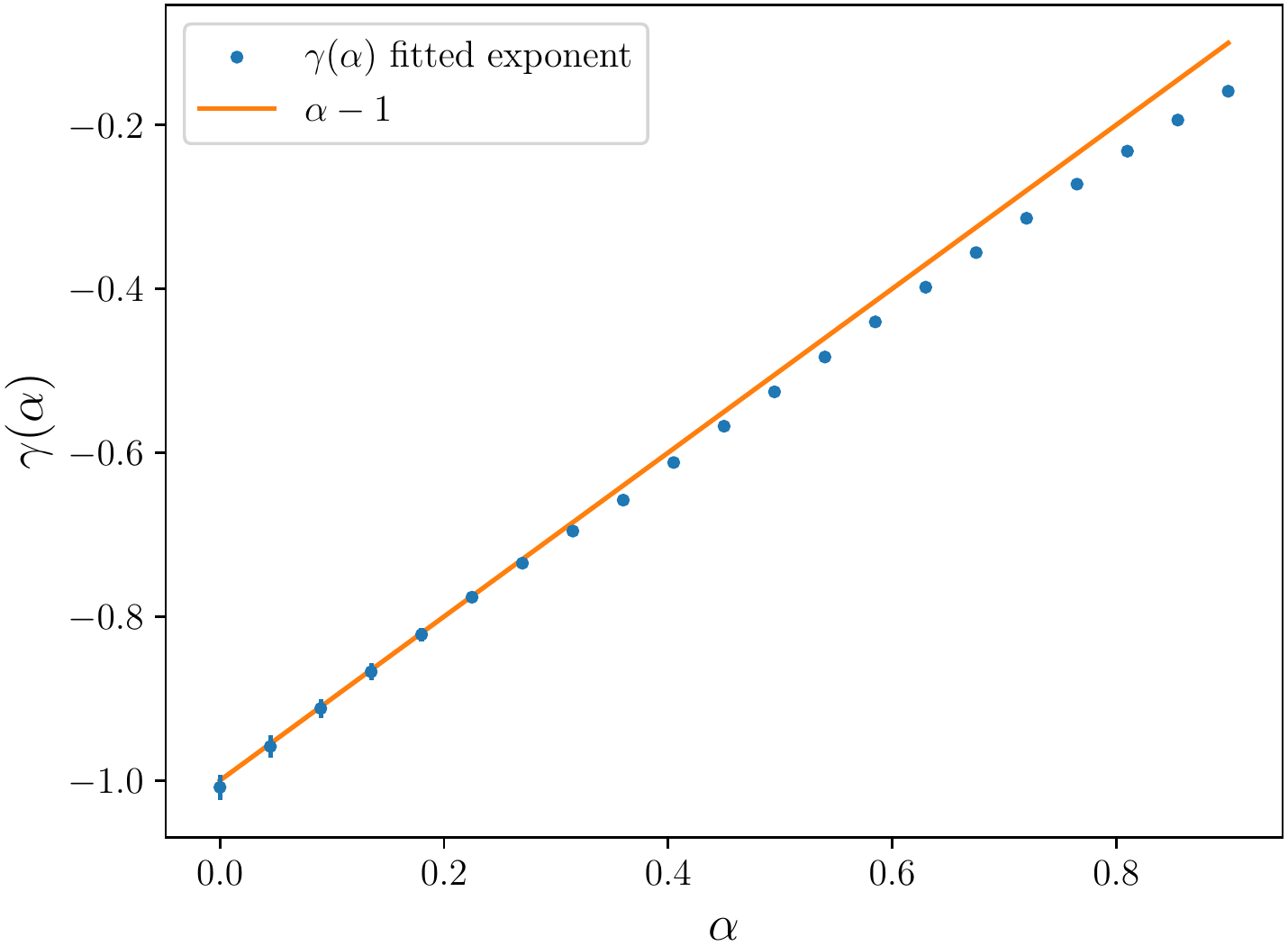}}
	\caption{The fitted exponent $\gamma$ in the signaling time scaling obtained from the exact summation bound in \cref{eq:exactbound} at $r=1$ (a) and $r=N/2$ (b) for $a\in [0,1]$. The error bars reflect the 95\% confidence intervals for the fit.}
        \label{Fig_al<D}
\end{figure*}
In \cref{Fig_al<D}, we plot the fitted exponent $\gamma(\al)$ as a function of $\al$. We observe that $\gamma(\al)$ scales approximately as $\al-1$ as long as $\al$ is not close to 1 for both $r=1$ and $r=N$, showing that both bounds lead to approximately the same scaling of signaling time in $N$.

When $\al\rightarrow 1$, $\gamma(\al)$ deviates from $\al-1$ noticeably. We attribute such deviation to finite-$N$ effects in our numerical evaluation of the exact summation bound. In particular, we notice that as $\al\rightarrow 1$, $\lambda$ (which plays an important role in both bounds) is not well-approximated by $N^{\al-1}$ for insufficiently large enough $N$. For such values of $N$, $\lambda$ is better approximated by $\log(N)$. 

To give further clarification, we perform a comparison of the two bounds exactly at $\al=1$, where we can exactly use $\log(N)$ in place of $N^{\al-1}$. The signaling time given by the analytical bound now scales as
\begin{equation}
	t_\text{si}(r,N) = \W{\frac{\log(r\log N)}{\log N}}.\label{eq:scalinga1}
\end{equation} 
We then fit the signaling time obtained from the exact summation bound at $\al=1$ using the above scaling function and find very good agreement between the two bounds. For example, at $r=1$ the signaling time from the exact summation bound can be fitted by the function $ a\:\log(N)^b\:\log\log(N)^c$ with $b=-1.0$ and $c=0.95$, which agrees with the scaling of $\log\log(N)/\log(N)$ provided by \cref{eq:scalinga1}. As a result, we expect the signaling time bounds given by both bounds to have the same scaling in $N$ when the system size is large enough for all $\al\le D$.

\makeatletter
\renewcommand\@biblabel[1]{[S#1]}
\makeatother
\bibliographystyle{apsrev4-1}
\bibliography{LRbib}